\documentclass[graybox]{svmult}

\usepackage{type1cm}        
\usepackage{float}
\usepackage{makeidx}         
\usepackage{graphicx}        
\usepackage{multicol}        
\usepackage[bottom]{footmisc}

\usepackage{newtxtext}       %
\usepackage[varvw]{newtxmath}       

\makeindex             

\def\beq{\begin{equation}}
\def\eeq{\end{equation}}
\def\ber{\begin{eqnarray}}
\def\eer{\end{eqnarray}}
\def\benu{\begin{enumerate}}
\def\eenu{\end{enumerate}}

\def\l{\left}
\def\r{\right}

\def\n{\nabla}

\def\f{\frac}
\def\mpl{m_{p}}

\def\nt{n_{_T}}

\def\Tre{T_{_{\rm re}}}

\def\wre{w_{_{\rm re}}}

\def\ng{n_{_{\rm GW}}}
\def\Og{\Omega_{_{\rm GW}}}

\def\sq{\lower.25ex\hbox{\large$\Box$}}
\def \lleq {\lower0.9ex\hbox{ $\buildrel < \over \sim$} ~}
\def \ggeq {\lower0.9ex\hbox{ $\buildrel > \over \sim$} ~}
\def\statei{\lbrace r,s \rbrace}

\def\atridot{\stackrel{...}{a}}

\def\apj{{Astroph.\@ J.\ }}

\def\prl{{Phys.\@ Rev.\@ Lett.\ }}
\def\prd{{Phys.\@ Rev.\@ D\ }}

\def\plb {{Phys.\@ Lett.\@ B\ }}

\def \jetpl {JETP Lett.\ }

\def\n {\noindent}

\begin{document}

\title*{Remembering Alexei Starobinsky -- the gentle giant of cosmology}

\author{Varun Sahni}

\institute{Inter-University Centre for Astronomy and Astrophysics,
Post Bag 4, Ganeshkhind, Pune 411~007, India.
 \email{varun@iucaa.in}
}

\maketitle

\abstract{I share fond memories of my former PhD advisor Alexei Starobinsky with whom I was closely associated for nearly 45 years. I reflect upon my early years in Moscow when I worked with him on my thesis, and touch upon the seminal work on inflation which he did during that period. Alexei visited India often and actively interacted with Indian scientists and students on
issues relating to inflation, large scale structure  and dark energy. This extensive collaboration, which lasted several decades, resulted in the publication of over a dozen important papers, several PhD's, and the development of the Statefinder and Om diagnostics, which I briefly discuss.
}



\section{Introduction}
\label{sec:intro}
I first met Alexei Starobinsky on March 29, 1979 which was a Thursday. I remember this day well because it also happened to be my birthday and I was
feeling quite happy and elated at having turned 23.
I had arrived in Moscow from New Delhi three years earlier in 1976 to pursue my higher studies at the physics department of Moscow State University. I was admitted into the second year of the five year integrated MSc course and began studying physics in the Russian language which, at first, was not an easy task for me. As my studies advanced I became increasingly fascinated by the field of cosmology and started attending the outstanding course taught by Prof. Yakov Zeldovich to students of Moscow state university\footnote{See my article \cite{zeldovich} for a summary of Zeldovich's many remarkable contributions to science.}. Zeldovich's academic style and personality attracted me like a magnet and I soon began working under his guidance for my `Kursovaya rabota' -- project work, which students needed to do during their intermediate years before before embarking on their MSc thesis. I thoroughly enjoyed working under Zeldovich, but when the time came for me to find a guide for my MSc and PhD Zeldovich gently informed me that he no longer guided students at the advanced level. He then generously offered to introduce me to any member of his talented team with whom I might like to pursue higher studies.

I immediately thought of Alexei Starobinsky, who, at the young age of 31, was widely considered to be the rising star of Soviet cosmology.
Zeldovich provided me with Alexei's contact details and so on March 29, 1979 I went to the Landau Institute on Sparrow street to meet my future guru and thesis supervisor. When I had asked Zeldovich earlier, on how I should address Dr Starobinsky, he had thought for a moment and then replied
``let us simply call him Alyosha''. Alyosha was the shortened nick-name with which everyone addressed Alexei Starobinsky and which was easier for me to pronounce than the more formal Alexei Alexandrovich according to Russian custom.

My association with Alyosha got off to a smooth start. He was very kind and considerate and began guiding me on how to carry out independent research in physics. At that point of time, in the 1970's, Alyosha was deeply involved in the exciting new subject area of `quantum field theory in curved space-time'. His 1971 paper with Zeldovich on the quantum mechanical production of particles in an anisotropic universe is widely considered to be a landmark paper in this field \cite{zeld-star71}. Zeldovich and Starobinsky demonstrated that a rapidly changing gravitational field could give rise to copious particle production in the early universe. This seminal work paved the way for the development of regularization and renormalization methods
which are essential for the extraction of a finite value of the vacuum energy momentum tensor, $\langle T_{\rm ik}\rangle$, from a formally divergent and infinite quantity. Of central importance was the fact that the backreaction of created matter on the background geometry, via the semi-classical Einstein equations $G_{\rm ik} = 8\pi G\langle T_{\rm ik}\rangle$, could rapidly isotropize an initially anisotropic universe \cite{lukash-star74}. This helped address
the famous observation by Collins and Hawking
\cite{collins-hawking} that
“the set of spatially homogeneous cosmological models which approach
isotropy at infinite times is of measure zero in the space of all spatially
homogeneous models…. It therefore seems that there is only a small set of initial
conditions that would give rise to universe models which would be isotropic to
within the observed limits at the present time.”
The work of Zeldovich and Starobinsky \cite{zeld-star71}
and Lukash and Starobinsky \cite{lukash-star74} showed that our universe
could arise from a much larger set of initial conditions, if quantum effects
were taken into account\footnote{Alexei revisited this issue in \cite{star83} where he showed that a positive cosmological constant could also
isotropize the universe. A similar result was obtained by Wald \cite{wald83} and later generalized to an inflationary universe by Moss and Sahni \cite{moss-sahni}. These results demonstrate that inflation can iron out anisotropies and small inhomogeneities thereby considerably enlarging the set of initial conditions from which our homogeneous and isotropic universe could arise.}.

\n
Decades later, Alexei once remarked to me, that his very first paper with Zeldovich \cite{zeld-star71}  -- written for his MSc -- was perhaps the best paper he had ever written !

For his PhD Alexei explored the possibility of quantum particle production from a rotating black hole \cite{star73}. His paper
“Amplification of waves reflected from a rotating black hole`` published in 1973 was pathbreaking and influenced Stephen Hawking's  famous work on black hole evaporation \cite{hawking74}.
As Hawking later reminiscenced \cite{hawking_book}
``In September 1973 while I was visiting Moscow, I discussed black holes with two leading Soviet experts, Yakov Zeldovich and Alexander Starobinsky. They convinced me that according to the QM uncertainty principle, rotating black holes should create and emit particles.''
In 1974 Hawking showed that quantum particle production is also a key feature of non-rotating black holes which shine via the famous `Hawking radiation' formula   $T = 1/8\pi M$ \cite{hawking74}.

I worked with Alexei from 1979 to 1985, initially for my MSc and then for my PhD. Most of Alexei's efforts during this period were directed towards developing inflationary cosmology.
Below I list some of his seminal results that emerged during 1979-1997.

\begin{itemize}
 \item 1979 Production of relic gravitational waves (GW) from Inflation \cite{star79}.

 \item 1980 First model of Inflation in the framework of $R + R^2$ gravity \cite{star80}.

 \item 1982 Determination of the primordial scalar power spectrum in Inflation \cite{star82}.

 \item 1985 Calculation of the tensor-to-scalar ratio $r$ in inflationary cosmology \cite{star85}.

 \item 1986 Developed the formalism of Stochastic Inflation \cite{star86,star-yokoyama}.

\item 1994-97 Developed the formalism of post-inflationary preheating
(with Kofman and Linde) \cite{kofman94,kofman96,kofman96a,kofman97}; also see \cite{yuri95}.

\end{itemize}

I should mention that during my years in Moscow office space was very limited, due to which I was obliged  to work from my hostel room while Alexei mostly worked from home. We agreed to meet on Thursdays at the Landau institute, provided I either had a nice result to show him, or -- which was more frequent --
I was fundamentally stuck in an intractable calculation. Occasionally I also visited Alexei at his stately home where his very kind wife Lyudmila made me some delicious Russian Borsch.

From the very outset Alexei struck me as a kind of {\em science wizard}.
No matter how mathematically complicated the problem, Alexei
could always guess the final answer even without performing any of the very tedious intermediate calculations (which were left to  me). During my visits to the Landau institute I was occasionally accompanied by Lev Kofman, who had started working under Alexei around the same time as me. Lev and I soon became close friends. Although Lev lived in the Estonian city of Tartu, he visited Moscow often and on these occasions used to stay in my small room in the Moscow university hostel. Lev and I had a great time together. We watched movies, had awesome discussions and performed difficult calculations seated side by side next to  my writing table (which also served as our dining table). Since living in the hostel required special permission, Lev's stay at my place was illegal and he was once hauled off by the police after a surprise raid. I rushed to the police station, located in the university basement, and begged and pleaded for the release of Lev. Luckily the police were lenient and let him go after a warning, after which all was forgotten and Lev began living in my room again. I defended my PhD in 1985 and my research papers included several collaborations with Lev. Alexei was the junior supervisor and Prof. Zeldovich the senior supervisor of my PhD thesis.

It is interesting that Alexei Starobinsky and Yakov Zeldovich had very contrasting science styles. Zeldovich was an intuitive physicist par excellence.
He had the gift of cutting to the very essence of complex physical phenomena and then explaining them to us students using very simple language. This was a key attraction of his bi-weekly seminar series at the Shternberg institute where he invited two experts to present their results, which they sometimes did in a very cumbersome manner. At the end of the presentation Zeldovich would bound up to the podium and lucidly explain to the audience the very gist of the problem being discussed. I could never quite figure out how he did this so spontaneously, since the seminar -- under the general title of `Unified Astrophysics Seminar' -- covered a wide range of topics ranging from stellar physics to quantum gravity.
Zeldovich's excellent lectures to students of Moscow state university were marked by the same style: rigorous, yet at the same time intuitive.
I believe that Zeldovich developed his unique scientific style because he was essentially self-taught, his formal education having ended very early -- in the 8th standard.

Alexei, on the other hand, was a formidable mathematical physicist.
His understanding of complicated physical problems was supported by a deep and nuanced knowledge of mathematics which he imparted to his students. Although I struggled at first during my PhD (partly since the text books were in Russian), those early years  of exposure to complicated mathematical concepts and calculations developed in me a deep sense of confidence that stood me in good stead in the years to come.

Alexei was the recipient of numerous international awards and honours including:
the Oscar Klein medal (2010), the Gruber cosmology prize (2013), the Kavli prize (2014) and the ICTP Dirac medal (2019). He was a fellow of several academic bodies including the Russian Academy of Science and the Indian National Science Academy.
Below I summarize some of the key scientific achievements of Alexei Starobinsky which were a major influence on my own work
in cosmology.

\section{Inflation}
\label{sec:inflation}
\subsection{\bf Relic gravitational waves from Inflation}
\label{sec:GW}

In 1979 Alexei examined an interesting cosmological scenario
in which the radiative epoch was preceded by a non-singular de Sitter universe\footnote{Gliner (1966-75) originally advocated the possibility of a metastable super-dense de Sitter stage in the early universe \cite{gliner}. Alexei was aware of Gliner's work when he wrote his GW paper in 1979 and cites it in \cite{star79,gliner1}. However Alexei was also looking for a physically better motivated scenario of a metastable early de Sitter stage which subsequently led him to develop his $R+R^2$ model of inflation in 1980 \cite{star80}.}. Alexei showed that
relic gravitational waves  were a generic prediction of such a scenario \cite{star79} -- subsequently called inflation \cite{guth81,linde82,alstein82}.
It is interesting that the amplitude of relic GWs
is sensitive to the value of the Hubble parameter
during inflation, while the GW spectrum encodes both the inflationary and post-inflationary
EOS of the universe \cite{star79,allen88,sahni90,giovani,sami2002,dany_18,dany_19,swagat}. Since GWs interact minimally with other forms of matter
they constitute one of the cleanest probes of physics of the very early universe.

\bigskip

In a homogeneous and isotropic universe
 massless scalar fields satisfy the Klein-Gordon equation
 \beq
 \left (\Box + \xi R\right )\Phi = 0
 \label{eq:KG}
 \eeq
where $R$ is the Ricci scalar and $\xi$ characterizes the coupling to gravity,
with $\xi = 1/6, 0$ corresponding to conformal and minimal coupling
respectively. In a spatially flat FRW universe
$\Phi = \phi_k(\tau) e^{-i{\bf k}.{\bf x}}$
where $\tau$ is the conformal time  $\tau = \int dt/a(t)$.
Performing the conformal transformation $\chi_k = a\phi_k$ and using $R = 6{\ddot a}/a$
leads to
\beq
{\ddot \chi}_k + \left \lbrack k^2 - V(\tau) \right\rbrack \chi_k = 0
\label{eq:KG2}
\eeq
where
\beq
V(\tau) = (1-6\xi)\frac{\ddot a}{a}\,.
\label{eq:barrier}
\eeq
For exponential inflation $a = \tau_0/\tau$ $(\vert\tau\vert < \vert\tau_0\vert)$, whereas $a \propto \tau, \tau^2$
in a post-inflationary radiation/matter dominated universe.

\begin{figure}[t]
\includegraphics[width=0.8\textwidth]{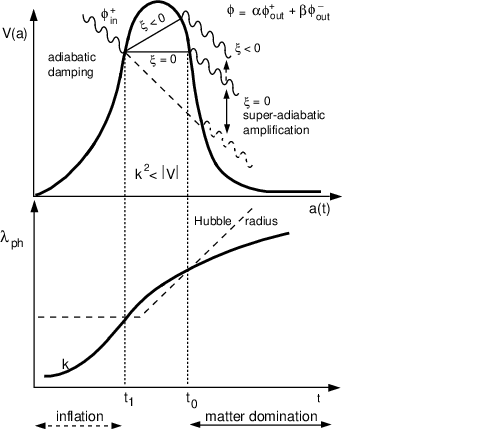}
\caption{{\bf GW production} is  illustrated by the
time-like ``potential barrier''
$V(a) = (1-6\xi) \frac{\ddot a}{a}$ in inflationary cosmology (top).
The evolution of the comoving Hubble
radius is shown below.  Short wavelength modes with $k^2 \gg |V|$ experience adiabatic damping (\ref{eq:damping}) and no particle production. Long wavelength modes, on the other hand, exit the Hubble radius at $t_1$ during inflation and
re-enter it during matter domination at $t_0$. The amplitude of these modes
remains frozen while they are outside the Hubble radius ($k^2 < |V|$).
As a result the gravitational
wave ($\xi = 0$) amplitude is  super-adiabatically amplified post-inflation. The amplification of fields with $\xi < 0$ can give rise to dark energy, as shown by Sahni and Habib \cite{sahni-habib}.}
\label{fig:barrier}
\end{figure}

Eqn (\ref{eq:KG2}) bears an interesting
resemblance to the one dimensional non-relativistic Schroedinger equation
\beq
\frac{d^2\Psi}{dr^2} + \left\lbrack E - V(r)\right\rbrack \Psi = 0~.
\label{eq:Schr}
\eeq
Comparing (\ref{eq:KG2}) and (\ref{eq:Schr}) one finds that
$V(\tau)$ in (\ref{eq:KG2})
plays the role of a `potential barrier in {\em time}', which is illustrated in
fig. \ref{fig:barrier}.

Two values of $\xi$ merit special mention.

For conformal coupling ($\xi = 1/6$), $V(\tau) = 0$, which implies $\phi(\tau) \propto e^{-ik\tau}/a$
and indicates that massless conformally coupled
scalars cannot be created in an expanding universe. This result applies to massless
neutrino's and photons, and for a while it was felt that all massless particles
may have this property \cite{parker68}. However, as shown by Grishchuk \cite{grish75}, gravitational
waves couple minimally ($\xi = 0$) to gravity and can therefore be created quantum mechanically
as the universe expands.

The two polarization states of the graviton satisfy
$$h_{\times,+} = \sqrt{8\pi G} \,\frac{\chi_k(\tau)}{a}\,e^{i{\bf k}{\bf x}}$$
where $\chi_k$ is determined from the wave equation (\ref{eq:KG2}) \&
(\ref{eq:barrier}).

Eqns. (\ref{eq:KG2}) \& (\ref{eq:barrier}) inform us that for short wavelength modes
with $k^2 \gg |V|$
\beq
\chi_k \propto e^{-ik\tau} ~~{\rm and} ~~ \phi_{\rm in}^+ = \frac{\chi_k}{a}
\propto \frac{e^{-ik\tau}}{a}\,,
\label{eq:damping}
\eeq
which corresponds to the adiabatic decay of positive frequency modes and no particle production.
However (\ref{eq:KG2}) \& (\ref{eq:barrier}) also reveal that long wavelength modes freeze out once they exit the Hubble radius since
\beq
\phi_k \simeq {\rm constant}\,, ~~{\rm for} ~~ k^2 < |V|\,.
\nonumber\\
\eeq
As a result these modes get super-adibatically amplified once the universe stops
inflating and enters a radiation or matter dominated stage, see fig. \ref{fig:barrier}. Thus
inflationary zero-point fluctuations get transformed into post-inflationary gravitational waves \cite{star79,grish75}.

\medskip

The dimensionless amplitude of a tensor GW mode is related to
the inflationary Hubble parameter ${H_k^{\rm inf}}$ at Hubble exit by
\beq
P_{\rm GW}(k) \equiv h^2_{\times,+}(k) \simeq \frac{1}{2\pi^2}\left (\frac{H_k^{\rm inf}}{\mpl}\right )^2\bigg\vert_{k=aH}~.
\label{eq:GW_Ph}
\eeq

The spectral density of stochastic GWs, defined in terms of the critical
density at the present epoch $\rho_{0c}$, is
\beq
\Og (f) \equiv \f{1}{\rho_{_{0c}}}\f{d\rho_{\rm GW}^0(f)}{d\log{f}}~.
\label{GW_Omega_def}
\eeq
During the post-inflationary epoch it is described by the following set of equations \cite{sahni90,dany_19,swagat}
\ber
\Og^{(\rm MD)}(f) &=& \frac{3}{8\pi^3}P_{\rm GW}(f)
\,\Omega_{0m}\left (\frac{f}{f_{h}}\right )^{-2},
~ f_h < f \leq f_{\rm eq}\label{eq:GW_spectrum_1a}\\
\Og^{(\rm RD)}(f) &=& \frac{1}{6\pi}P_{\rm GW}(f)\,\Omega_{0r},
~ f_{\rm eq} < f \leq f_{\rm re}\label{eq:GW_spectrum_1b}\\
\Og^{(\rm re)} (f) &=& \Og^{(\rm RD)}
\left (\frac{f}{f_{\rm re}}\right )^{2\left (\frac{w-1/3}{w+1/3}\right )},
~ f_{\rm re} < f \leq f_{\rm e}\,.
\label{eq:GW_spectrum_1c}
\eer
The superscripts `MD', 'RD', `re' in $\Og$ refer to the matter dominated epoch, radiative epoch and the epoch of reheating respectively, while
$f_h$, $f_{\rm eq}$, $f_{\rm re}$, $f_e$ refer  to the present day frequency of relic GWs corresponding to tensor modes that became sub-Hubble at: the present epoch ($f_h$),
the epoch of matter-radiation equality ($f_{\rm eq}$), the commencement  of the radiative epoch ($f_{\rm re}$) and at the end of inflation ($f_e$). Note that $f_{\rm re}>f_{\rm BBN} \simeq 10^{-11}~{\rm Hz}$ in order to
satisfy the BBN bound on the reheating temperature $\Tre$. In (\ref{eq:GW_spectrum_1c}) $w$ is the post-inflationary equation of state (EOS) during the epoch of reheating.

The inflationary power spectrum $P_{\rm GW}(k)$ can be written as
\beq
P_{\rm GW}(k) = P_{\rm GW}(k_*)\left (\frac{k}{k_*}\right )^{\nt}~,
\label{eq:GW_Ph1}
\eeq
where the tensor power  at the CMB pivot scale $k_* = 0.05 ~{\rm Mpc}^{-1}$ is  given in terms of the scalar power spectrum $A_{_S}$ by
\beq
P_{\rm GW}(k_*) = \frac{1}{4}\, r \, A_{_S} =r\times 5.25 \times 10^{-10}~,
\label{eq:GW_Ah}
\eeq
and the tensor tilt is found to be
\beq
n_{_{\rm GW}}\equiv n_T = \frac{d\log{P_{\rm GW}(k)}}{d\log{k}} = -\f{r}{8}~,
\label{eq:GW_nt}
\eeq
which satisfies the consistency relation; see \cite{swagat} for details.

\begin{figure}[b]
\centering
\includegraphics[width=0.85\textwidth]{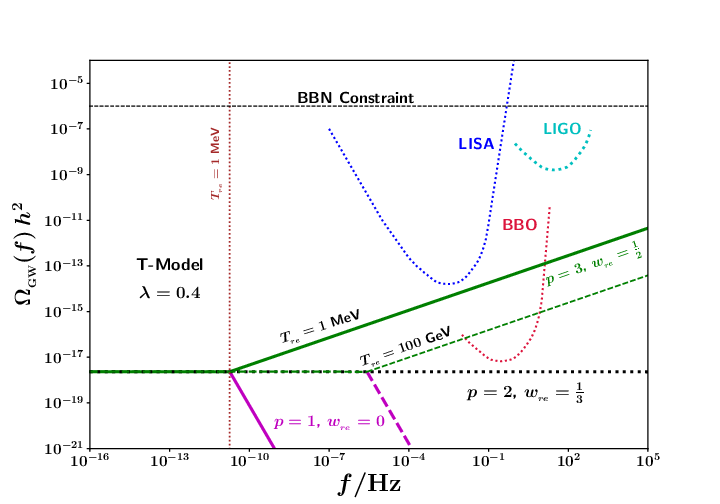}
\caption{The spectrum of relic gravitational waves is plotted for the T-model $\alpha$-attractor potential \cite{Kallosh-Linde}
$V(\phi) = V_0 \tanh^{2p} (\frac{\lambda\phi}{m_p})$ for $\lambda=0.4$ assuming that reheating takes place perturbatively. The dotted black curve corresponds to $p=2$ for which the post-inflationary EOS is radiation-like, with $\wre=\l<w_{_{\phi}}\r>=1/3$. The solid and dashed green curves corresponds to reheating temperatures $\Tre = 1~{\rm MeV}$ and $100~{\rm GeV}$ respectively, for the case when $p=3$ where the  post-inflationary EOS is stiff with $\wre=\l<w_{_{\phi}}\r>=1/2$. We infer that  blue tiled relic GWs can be detected by the future GWs observatories such as the Big Bang Observatory (BBO) \cite{BBO}, for a range of $\Tre$.  A matter-like post inflationary EOS
with $\wre=\l<w_{_{\phi}}\r>=0$ corresponds to $p=1$, and is shown in red (solid, dashed). One sees that the red tilted GW amplitude is strongly suppressed in this case. From \cite{swagat}.}
\label{fig:GW_Tmodel}
\end{figure}

The GW spectrum, $\Og(k)$,
and spectral index $\ng = \frac{d\log{\Omega_{g}}}{d\log{k}}$,
therefore serve as
key probes of physical processes occurring both during and after inflation.
As originally shown in \cite{star79,allen88,sahni90}, and outlined
in (\ref{eq:GW_spectrum_1a}) - (\ref{eq:GW_spectrum_1c}),
the spectrum of relic gravitational radiation is exceedingly sensitive to the {\bf post-inflationary
EOS $w$}.
From (\ref{eq:GW_spectrum_1c}) one finds that the GW spectrum on small scales
can have a red tilt (for $w < 1/3$), a blue tilt (for $w > 1/3$) or be scale free
(if $w = 1/3$). The GW spectrum therefore provides us with a powerful signature of
physics in the early universe. Some consequences of blue tilted and red tilted GW spectra are shown
in figure \ref{fig:GW_Tmodel} and a more comprehensive discussion of this issue is
given in \cite{swagat}.
Concerning the EOS $w = \wre$ during the post-inflationary epoch of reheating, it is important to keep in mind the issues summarized in the next section.

\subsubsection{Reheating}
\label{sec:reheating}

A key feature of inflationary cosmology is that it enables the universe to reheat by transferring
the energy localized in the inflaton to the matter/radiative degrees of freedom present in the universe.

In potentials possessing a minimum,
reheating can occur in two distinct ways: (i) perturbatively (slowly),
(ii) rapidly -- via a parametric resonance.
Which of these two ways is realized depends upon the nature of the coupling between the inflaton
and bosons/fermions.
In the case of perturbative reheating, the scalar field oscillates
for a very long time gradually releasing its energy into matter/radiation.
In this case the EOS during the oscillatory regime, $\langle w_{_{\phi}}\rangle$,  is determined primarily by the shape of the
inflaton potential near its minimum value, about which the inflaton oscillates. For $V \propto \phi^{2p}$, $\langle w_{_{\phi}}\rangle = \frac{p-1}{p+1}$.
Perturbative reheating is expected to occur if the inflaton decays primarily into fermions
(which then decay into standard model fields), its decay into bosons being strongly suppressed
\cite{kofman96a}.

However if the inflaton decays into bosons, $\chi$, through the coupling $g^2\phi^2\chi^2$
with $g \gg 10^{-3}$, then oscillations of $\phi$ can lead to a parametric resonance during which
quanta of the field $\chi$ are produced in copious amounts. This stage is usually referred to as
preheating and was studied in great detail by Alexei Starobinsky and his colleagues in \cite{kofman96a,kofman94,yuri95,kofman96,kofman97}.
The backreaction of $\chi$ on $\phi$ ends the resonance and the subsequent decay of excitations
of the $\phi$ and $\chi$ fields into standard model (SM) fields gives rise to reheating
and the subsequent thermalization of the universe at a temperature $\Tre$.
The duration of the pre-radiative epoch, which includes the end of inflation, the parametric
resonance, the decay of the inflaton into bosons ($\phi \to \chi\,\chi$) and fermions
($\phi \to \psi\,\psi$) and thermalization can be quite long,
and it is convenient to encode its physics by means of an effective EOS parameter $\wre$.
Since $\wre$ influences the spectrum of relic gravitational waves, observations of the GW
spectrum can shed light on the complex, non-linear physics
which operates during the reheating epoch (this sub-section has been adapted from
\cite{swagat}).

\subsection{$R + R^2$ Inflation}
\label{sec:r2inflation}

In 1980 Alexei suggested an interesting model of inflation which is
commonly referred to as `Starobinsky inflation' \cite{star80}. It is based on the Jordan frame action\footnote{The quadratic $R^2$ term can appear as a one loop quantum correction to GR \cite{birrel-davies}. Eqn (\ref{eqn:star_action}) bears an interesting analogy with the quantum (Euler-Heizenberg) corrections to classical electrodynamics. The classical electromagnetic Lagrangian ${\cal L}_{\rm cl} = -\frac{1}{4}F_{\mu\nu}F^{\mu\nu} \equiv \frac{1}{2}({\bf E}^2 - {\bf B}^2)$, in the presence of one-loop quantum corrections becomes
${\cal L} = {\cal L}_{\rm cl} + {\cal L}_{\rm Q}$
where the quadratic quantum correction
\beq
{\cal L}_{\rm Q} = \frac{\alpha^2}{90 m_e^4}\left \lbrack \left (F_{\mu\mu}F^{\mu\nu}\right )^2 + \frac{7}{4}\left (F_{\mu\nu}{\tilde F}^{\mu\nu}\right )^2\right\rbrack \equiv \frac{2\alpha^2}{45m_e^4}\left\lbrack\left ({\bf E}^2 - {\bf B^2}\right )^2 + 7\left ({\bf E}.{\bf B}\right )^2\right\rbrack
\eeq
leads to third order equations and non-linear electrodynamics.
($\alpha$ is the fine structure constant.)
}

\begin{equation}
S=\int d^{4}x\sqrt{-g} \frac{m_{p}^{2}}{2}\bigg [R+\frac{1}{6m^{2}}R^{2}  \bigg ]
\label{eqn:star_action}
\end{equation}
where $m$ is a mass parameter and $m_p = 1/\sqrt{8\pi G}$. 
Since
\beq
\frac{\delta}{\delta g^{\rm ik}}\int d^4x \sqrt{-g}\frac{R^2}{m^2} = 2 g_{\rm ik} \Box R + \cdots
\eeq
this action leads to fourth order gravity, with two orders
being encoded in the $\Box$ operator and another two in the Ricci scalar $R$.

\begin{figure}[b]
\centering
\includegraphics[width=0.48\textwidth]{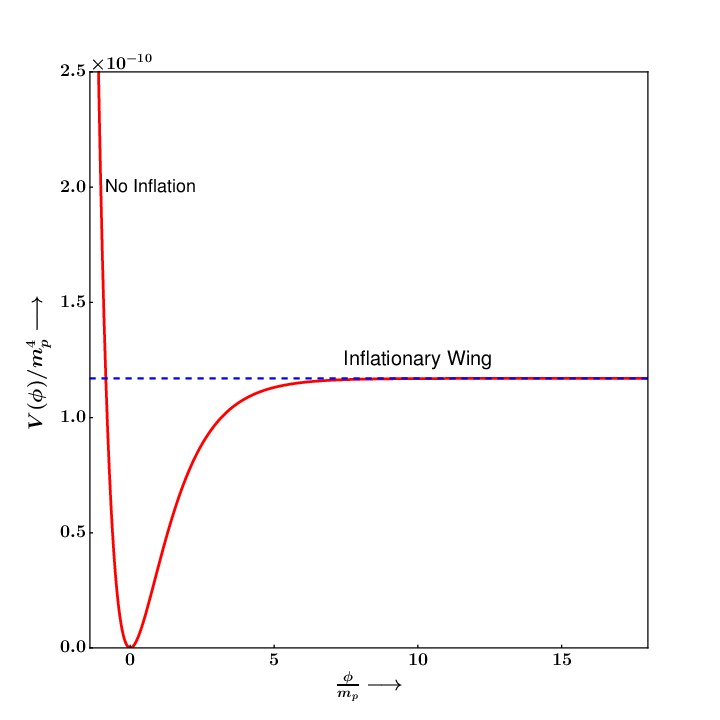}
\includegraphics[width=0.48\textwidth]{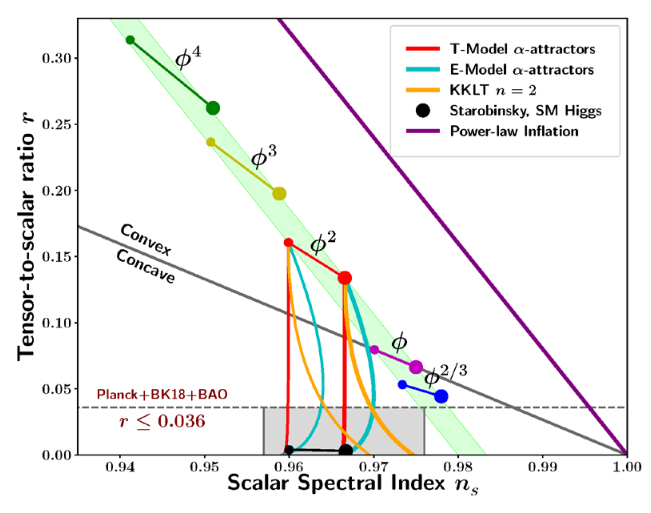}
\caption{{\bf{Left:}} The effective potential in Starobinsky Inflation, (\ref{eqn:star_pot}), is plotted  in units of $m_{p}^{4}$.
The potential is asymmetric about the origin and has a steep left wing and  plateau-like right wing.
Inflation occurs along the flat right wing,
the steep left wing is unable to sustain inflation.
{\bf {Right:}}The tensor-to-scalar ratio $r$ and the scalar spectral index $n_s$ from the BICEP-Keck data \cite{BICEP-Keck} are shown
for several inflationary models. While most early models of
inflation are either strongly constrained or ruled out by CMB data, the Starobinsky model provides an excellent fit. From \cite{swagat}.}
\label{fig:star_pot}
\end{figure}

It is sometimes convenient to write the action (\ref{eqn:star_action}) in
the Einstein frame where it has the form
\begin{equation}
S_{E}=\int d^{4}x\sqrt{-g} \bigg [ \frac{m_{p}^{2}}{2}\hat{R}-\frac{1}{2} \hat{g}^{\mu\nu}\partial_{\mu}\phi\partial_{\nu}\phi-V(\phi) \bigg ]
\label{eqn:star_action_einstein}
\end{equation}
where the inflaton potential is \cite{whitt84,maeda88}
\begin{equation}
V(\phi)=\frac{3}{4}m^{2}m_{p}^{2}\Big(1-e^{-2\frac{\phi}{\sqrt{6}m_{p}}}\Big)^{2}\,.
\label{eqn:star_pot}
\end{equation}

\n
The CMB determined amplitude of scalar fluctuations sets
$m=1.25\times 10^{-5}m_{p}$  \cite{gorbunov13}.
The potential (\ref{eqn:star_pot}) is shown in figure \ref{fig:star_pot} (left) and current CMB constraints on $R+R^2$ inflation are shown in the right panel.
As shown in that figure the potential for Starobinsky inflation is asymmetric about the origin.
The right wing of the potential is flat and has the  same functional form as the Higgs inflation potential in the Einstein frame. However the left wing of $V(\phi)$ is very steep. The slow-roll parameter for this potential is given by
$$\epsilon = \frac{m_{p}^2}{2}\left (\frac{V'}{V}\right )^2=\frac{4}{3}\l[\frac{e^{-\frac{2}{\sqrt{6}}\frac{\phi}{m_{p}}}}{1-e^{-\frac{2}{\sqrt{6}}\frac{\phi}{m_{p}}}}\r]^{2}\,.$$
 Inflation ends when
$\epsilon\geq 1$ which corresponds to
$\phi\geq 0.937$ and indicates that
no inflation can arise from the steep left wing of the potential
for which $\phi < 0$ \cite{mishra-inflation}.

It is interesting that in formulating his $R + R^2$ model of inflation Alexei was attracted by the fact that de Sitter space can arise as a self-consistent solution to the Einstein equations with quantum corrections in the RHS
\beq
G_{\rm ik} = 8\pi G \langle T_{\rm ik}\rangle\,,
\label{eq:one loop}
\eeq
a fact that was originally noted by Dowker and Critcheley in 1976 \cite{dowker}. (The $R^2$ term in the $R + R^2$ action generates a
higher order $\Box R$ term in $\langle T_{\rm ik}\rangle$.) However Alexei went a step further and showed that de Sitter space -- induced by its own vacuum polarization -- was unstable and made a transition to the FRW universe. The Starobinsky model therefore successfully  accommodated acceleration in the past (inflation) with deceleration at present. Self-consistent solutions to (\ref{eq:one loop}) were further explored in \cite{sahni-kofman} and references therein.\footnote{In his last conversation with me Alexei mentioned that he no longer believed that his $R + R^2$ model arose because of quantum corrections of the form (\ref{eq:one loop}) but might have a different origin.}

\subsection{Stochastic Inflation}
\label{sec:stochastic}

In his landmark 1986 paper \cite{star86}, Alexei showed that
``The dynamics of a large-scale quasi-homogeneous scalar field
producing the de Sitter (inflationary) stage in the early universe is strongly
affected by small-scale quantum fluctuations of the same scalar field and,
in this way, becomes stochastic. The evolution of the corresponding large
scale space-time metric follows that of the scalar field and is stochastic also.''
Alexei's formalism for stochastic inflation was very influential and paved the way for further important developments including {\em eternal inflation} and the {\em multiverse} \cite{star-yokoyama,linde86,linde08}; see \cite{woodard} for a recent review of stochasic inflation dedicated to the memory of Alexei Starobinsky.

There is an interesting parallel between stochastic inflation and classical diffusion
(Brownian motion) which allows one to formulate a diffusion equation for the
inflaton probability distribution $P(\phi,t)$ akin to the Fokker-Planck equation
\beq
\frac{\partial P(\phi,t)}{\partial t} = D\frac{\partial^2P(\phi,t)}{\partial\phi^2}
\eeq
where $D = H^3/8\pi^2$ is the diffusion coefficient. Solving this equation results in a Gaussian probability distribution for $P(\phi,t)$
\beq
P(\phi,t) = \frac{1}{\sqrt{2\pi}\sigma(t)} \exp\left (-\frac{\phi^2}{2\sigma^2(t)}\right )
\label{eq:gauss}
\eeq
where $\sigma^2 = \langle\phi^2\rangle $ is the quantum dispersion.
In de Sitter space $\langle\phi^2\rangle = \frac{H^3}{4\pi^2}(t-t_0)$
where $t_0$ marks the start of inflation \cite{vilenkin-ford}.
The growth in $\sigma^2$ broadens  the Gaussian distribution (\ref{eq:gauss}) with time and arises because of the quantum diffusion of the inflaton.

In the presence of a potential $V(\phi)$ the Fokker-Planck equation gets modified to
\beq
\frac{\partial P(\phi)}{\partial \log a} = \frac{1}{8\pi G}\frac{\partial}{\partial\phi}\left( \frac{V'}{V} P \right) + \frac{\partial}{\partial\phi} \left( \sqrt{\Delta}\frac{\partial}{\partial\phi}\left (\sqrt{\Delta} P\right)\right ), ~~~ \Delta = \frac{H^2}{8\pi^2}~.
\eeq 
This equation has the solution
\beq
P(\phi,t) \propto \exp\left\lbrack - \frac{\left (\phi-\phi_{\rm cl}(t)\right )^2}{2\sigma^2(t)}\right \rbrack
\eeq
$P(\phi,t)$ describes a Gaussian which is peaked about the classical trajectory $\phi_{\rm cl}$
\beq
{\ddot\phi}_{\rm cl} + 3H{\dot\phi}_{\rm cl} + V' = 0 .
\label{eq:phi-cl}
\eeq
Since $\sigma^2 \propto (t-t_0)$, the Gaussian broadens with time implying the quantum diffusion of the inflaton $\phi$ about its classical trajectory.

\begin{figure}[b]
\centering
\includegraphics[width=0.8\textwidth]{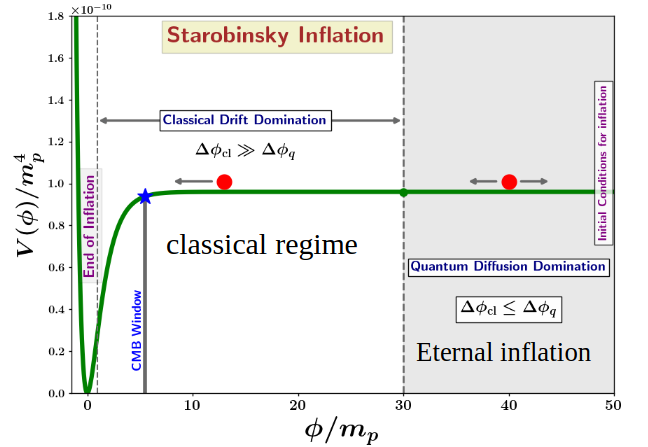}
\caption{An illustration of stochastic Starobinsky inflation (\ref{eqn:star_pot}).
}
\label{fig:stochastic}
\end{figure}

In the presence of quantum noise, and in the slow roll limit when
${\ddot\phi}_{\rm cl} \simeq 0$, the classical equation of motion
(\ref{eq:phi-cl}) gets transformed into the Langevan equation for the inflaton
$\phi = \phi_{\rm cl} + \Delta\phi_q$
\beq
\frac{d\phi}{d\log a} = -\frac{V'}{3H^2} + \frac{H}{2\pi}f(\phi,\log a)
\eeq
where $f$ is a stochastic force which describes random kicks that
are imparted to $\phi$ as small wavelength fluctuations of quantum origin  ($\Delta\phi_q$) grow in scale and cross the Hubble radius.
If the quantum noise dominates over the classical drift then
the universe can inflate eternally in some regions of space \cite{linde86},
as shown in fig. \ref{fig:stochastic}.

\section{Dark Energy}
\label{sec:DE}

After completing my PhD in Moscow in 1985, I went to the UK and to Canada for post-doctoral work. I finally returned to India in 1991 to join a newly established institute of
astrophysics in Pune called IUCAA. When I arrived the institutional campus was just being built and we were all housed in temporary quarters. What struck me immediately upon arrival were the large
number of old and beautiful Banyan trees on  campus and the copious abundance of snakes and other wildlife including peacocks, mongoose families and even scorpions (I was bitten by one).

At the time of my arrival the total faculty strength at IUCAA was only 6 and so we were very keen to have distinguished scientists
visit our institute and interact with members of the Indian scientific community. I applied for an Indo-Russian collaboration
program which allowed Alexei to visit India. He loved the ambience of IUCAA and, over the course of several visits, collaborated
with a large number of IUCAA students, visitors and faculty including Tarun Souradeep, Dipak Munshi, Tarun Saini, Ujjaini Alam, Arman Shafieloo, Swagat Mishra, M. Sami, Minu Joy and Somak Raychaudhury.

\subsection{Reconstructing dark energy}
\label{sec:recon}

The discovery of dark energy in 1998 \cite{DE} took almost everyone by surprise. Although the existence of a cosmological constant $\Lambda$ had been seriously discussed for several decades, its theoretical value was presumed to
be very large (formally infinite) and so the fact that $\Lambda$ may be small enough to describe the currently accelerating universe seemed unlikely.
However a small cosmological constant was anticipated in several key early papers. For instance in 1967  Zeldovich suggested that a small value of $\Lambda$ could be associated with fundmamental physics \cite{zel68} . Then in 1985 Alexei Starobinsky and Lev Kofman \cite{kofman-star-CMB}
performed the first ever calculation of large scale anisotropies
in the CMB in $\Lambda$CDM cosmology (the Sachs-Wolfe effect). In 1992 I had written a paper with Hume Feldman and Albert Stebbins on a universe
which {\em loiters} because of the presence of dark energy \cite{loiter},
and in 1998, at  the time of the SNIa discovery, I was working on a paper
with Salman Habib in which we showed that quantum effects during inflation could generate a small cosmological constant \cite{sahni-habib}. Because of this  early work, both Alexei and I were psychologically prepared for $\Lambda$ and began working on dark energy (DE) soon after its discovery. (See \cite{peebles-ratra,ostriker-stein} for other important early work on DE and \cite{de_rev} for reviews.)

The SNIa discovery of an accelerating universe led to many interesting papers being written which explained cosmic acceleration using new theoretical models.
Alexei and I followed a parallel path and developed model independent methods which could reconstruct the nature of DE directly from observations using model agnostic techniques \cite{saini00,ss06}.

The idea behind DE reconstruction is simple.
For standard candles (SNIa) with luminosity $L$, the flux received by an
observer is 
\beq
{\cal F} = \frac{L}{4\pi d_L^2}\,,
\eeq
where
\beq
d_L(z) = (1+z)\int_0^z\frac{dz'}{H(z')}\,
\label{eq:lum-dis}
\eeq
is the luminosity distance. From (\ref{eq:lum-dis})
one can obtain the expansion history
\beq
H(z) = \left \lbrack \frac{d}{dz}
\left (\frac{d_L}{1+z}\right)\right\rbrack^{-1}
\label{eq:Hz}
\eeq
and the equation of state of DE
\beq\label{eq:state}
w(z) = \frac{\frac{2x}{3}\frac{d \log{H}}{dx} - 1}{1 - \left (\frac{H_0}{H}\right )^2 \Omega_{0m} x^3}\,,  ~~ x = 1+z\,.
\eeq

Differentiating a noisy observable
such as the luminosity distance $d_L(z)$
increases the noise,
therefore one must smoothen $d_L(z)$ or $H(z)$
before differentiating in order to obtain $w(z)$.
This can be done either by
approximating $d_L(z)$ or $H(z)$
by an ansatz, or by smoothing
the data directly.
There are therefore two main approaches to DE reconstruction: parametric and non-parametric \cite{ss06}.

\begin{itemize}

\item
In the parametric approach one approximates a key cosmological quantity such as the luminosity distance $d_L(z)$ (or angular size distance $d_A$), the expansion history $H(z)$ or the DE equation of state, $w(z)$, using a simple ansatz such as
$w(z) = w_0 + w_1 \frac{z}{1+z}$ \cite{cpl}. The free parameters $w_0, w_1$ are then obtained by matching with observations.

\item
Non-parametric reconstruction is based on the observation that
one can obtain a smoothed quantity
$D^S({\bf x})$ from a fluctuating `raw' quantity,
$D({\bf x'})$, using a low pass filter $F$ having a
characteristic scale $R_f$ \cite{coles}
\beq
D^S({\bf x}, R_f) = \int D({\bf x'}) F(|{\bf x}-{\bf x'}|;R_f)~d{\bf x'}~.
\eeq
Commonly used filters include the Gaussian
filter $F_{\rm G} \propto \exp(-|{\bf x}-{\bf x'}|^2/2R_{\rm G}^2)$.
In studies of large scale structure $D$ is the density fluctuation $\delta({\bf x})$, whereas for
DE reconstruction $D$ could be either
$H(z), d_L(z), d_A(z)$, etc.

Since we wish to smooth the noise and not the signal we proceed as follows \cite{arman}
\ber
\label{eq:bg}
\ln d_L(z,\Delta)^{\rm s}=\ln
\ d_L(z)^g+N(z) \sum_i \left [ \ln d_L(z_i)- \ln
\ d_L(z_i)^g \right] &&\nonumber\\
{\large \times} \ {\rm exp} \left [- \frac{\ln^2 \left
( \frac{1+z_i}{1+z} \right ) }{2 \Delta^2} \right ], &&\\
N(z)^{-1}=\sum_i {\rm exp} \left
[- \frac{\ln^2 \left ( \frac{1+z_i}{1+z} \right ) }{2 \Delta^2}
\right ]~. \hspace{2.8cm}&&\nonumber
\eer
Here, $\ln d_L(z,\Delta)^{\rm s}$ is the smoothed luminosity distance
at redshift $z$ and $N(z)$ is a normalization
parameter.
This procedure is to be used iteratively with $\Lambda$CDM being the
first `guess' model for $d_L = d_L(z)^g$.
Convergence takes place quite rapidly
within only a few iterations.
Using $d_L(z) \equiv d_L^s(z)\big\vert_{\rm final}$ one can obtain the expansion history (\ref{eq:Hz})
and the EOS of DE (\ref{eq:state}).
A parallel approach involving cosmological reconstruction using Gaussian processes is discussed in \cite{arman-Gauss}.

\end{itemize}

\subsection{Null tests of $\Lambda$CDM}

\begin{figure}[htb]
\centering
\includegraphics[width=0.7\textwidth]{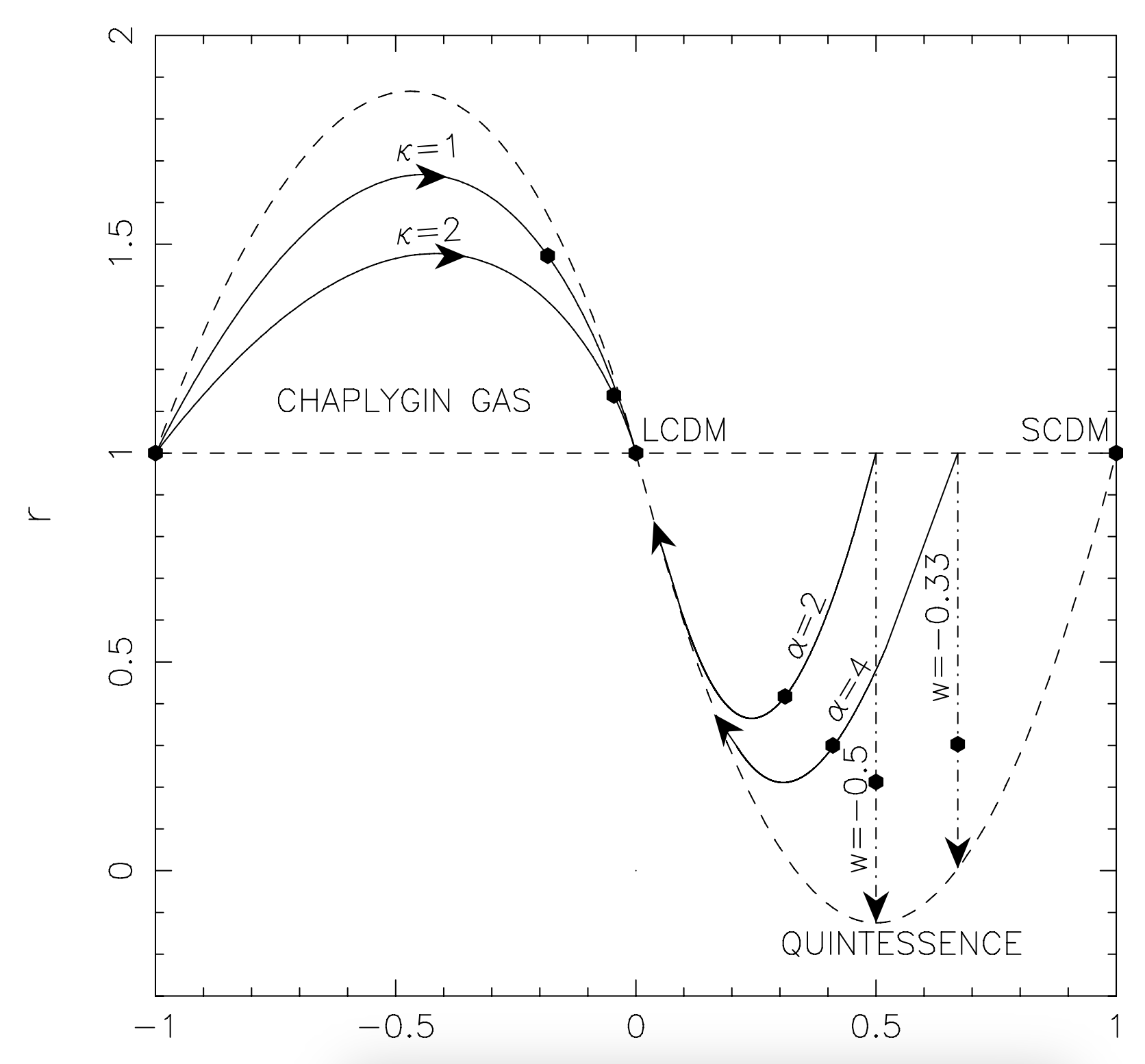}
\caption{The time evolution of the statefinder pair
$\lbrace r,s \rbrace$ for quintessence models with $V=V_0/\phi^{\alpha}$,
DE with a constant EOS $w$, and the Chaplygin gas.
Quintessence models lie to the right of the $\Lambda$CDM fixed point
($r=1,s=0$) while the Chaplygin gas lies to the left.
Both models approach the $\Lambda$CDM limit of $r=1,s=0$ in the future. For
the Chaplygin gas, $\kappa$ is the ratio between the matter density and the
density of the Chaplygin gas at early times.
The dashed curve in the lower right is the envelope of all
quintessence models, while the dashed curve in the upper left is the
envelope of Chaplygin gas models. In DE models with a constant EOS
$s$ remains
fixed at $1+w$ while $r$ declines asymptotically to
$1+\frac{9}{2}w(1+w)$.  From \cite{alam}.}
\label{statefinder}
\end{figure}

Of all Dark Energy models the cosmological constant, $\Lambda$, is singled
out by its elegance and simplicity since $
T_i^k = \Lambda\delta_i^k\,.$
So, as a first step, its logical to find tests which could falsify
the {\em cosmological constant hypothesis}. These include the Statefinder and $Om$ diagnostics.

\subsubsection{The Statefinder diagnostic of dark energy}
\label{sec:statefinder}

A general form for the expansion
factor of the Universe is given by the Taylor series expansion
\beq\label{eq:taylor0}
a(t) = a(t_0) + {\dot a}\big\vert_0 (t-t_0) +
\frac{\ddot{a}\big\vert_0}{2} (t-t_0)^2 +
\frac{\atridot\big\vert_0}{6} (t-t_0)^3 + ...~.
\eeq
In 1970 Alan Sandage described observational cosmology as being
``a search for two numbers''
\beq
H_0 = \left (\frac{\dot a}{a}\right )_0 ~~{\rm and} ~~ q_0 = -\left (\frac{\ddot a}{aH^2}\right )_0\,.
\eeq

In 2002 Alexei and I, together with
my students Ujjaini Alam and Tarun Saini,
showed that a new diagnostic of dark energy
-- Statefinder -- could be constructed using the third
derivative of the expansion factor \cite{sssa02}.
The statefinder pair $\statei$, defines two new cosmological
parameters (in addition to $H$ and $q$):
\ber\label{eq:statefinder1}
r &\equiv& \frac{\atridot}{a H^3} = 1 + \frac{9w}{2} \Omega_{_{\rm DE}}(1+w) -
\frac{3}{2}  \Omega_{_{\rm DE}}\frac{\dot{w}}{H} \,\,,\\
s &\equiv& \frac{r-1}{3(q-1/2)} = 1+w-\frac{1}{3} \frac{\dot{w}}{wH}\,\,.
\eer

A key property of the Statefinder is  that $\statei = \lbrace
1,0\rbrace$ is a {\bf fixed point} for $\Lambda$CDM.

The statefinder therefore provides us with a {\em fingerprint} of DE
since for all DE models excluding $\Lambda$CDM the trajectory of $\lbrace r,s\rbrace$
evolves with time.
In figure \ref{statefinder} we show
the behaviour of the statefinder pair $\lbrace r,s \rbrace$ for
a number of DE models including
\begin{enumerate}
\item

 Quintessence models defined by ${\cal L} = \frac{1}{2}{\dot\phi}^2 - V(\phi)$  and $V = V_0/\phi^\alpha$.
\item
 DE with a constant EOS $p = w\rho$.

\item
The Chaplygin gas with ${\cal L} = V_0\sqrt{1-\phi_{,\mu}\phi^{,\mu}}$ \, and \, $p_c = -A/\rho_c$.

\end{enumerate}
Note that null tests
 of $\Lambda$CDM can
also be constructed using higher derivatives of $a(t)$, which results in a
'Statefinder hierarchy' of null tests \cite{maryam}.

\subsubsection{The $Om$ diagnostic}
\label{sec:Om}

The $Om$ diagnostic is defined very simply as \cite{Om}
\beq
Om(z) \equiv \frac{h^2(z)-1}{(1+z)^3-1}\,,
\label{eq:om}
\eeq
where $h(z) = H(z)/H_0$.
The possibility of using $Om$ as a {\em null diagnostic} follows from
the fact that for $\Lambda$CDM
\beq
Om(z) = \Omega_{0m}~.
\label{eq:LambdaOm}
\eeq
In other words, the value of $Om$ is {\em redshift independent} for
the cosmological constant, while for other models of dark energy $Om(z)$
is {\underline {redshift dependent}}. Interestingly,
$Om(z) > \Omega_{0m}$ in quintessence models while
$Om(z) < \Omega_{0m}$ in phantom DE.

This important property of $Om$ can be used to considerable advantage by
defining the {\em difference diagnostic}
\beq
Om_{\rm diff}(z_1,z_2) := Om(z_1) - Om(z_2)\,.
\label{eq:om2}
\eeq
From (\ref{eq:LambdaOm}) one immediately finds that for the cosmological
constant
\beq
Om_{\rm diff}(z_1,z_2) = 0.
\label{eq:defn1}
\eeq
A departure of $Om_{\rm diff}$ from zero
therefore serves as a `smoking gun'
test for $\Lambda$CDM.
Since $Om$ depends only on the Hubble parameter and not its derivatives it may be easier to determine from (noisy) observations than either
$q(z)$ or $w(z)$. From (\ref{eq:om}) one finds that the $Om$ diagnostic does not depend upon the value of $\Omega_{0m}$ and is therefore insensitive to observational uncertainties in that quantity, unlike $w(z)$ which explicitly depends upon
$\Omega_{0m}$ through (\ref{eq:state}).

One can also define the two-point $Om$ diagnostic \cite{om3}
\beq
Om(z_2;z_1) = \frac{h^2(z_2)-h^2(z_1)}{(1+z_2)^3 - (1+z_1)^3}\,, ~~h(z) = H(z)/H_0\,,
\label{eq:om1}
\eeq
where $Om(z;0) = Om(z)$ and $Om(z_2;z_1) = \Omega_{0m}$ for $\Lambda$CDM.
Multiplying both sides of (\ref{eq:om1}) by $h_{100}^2$ where
 $h_{100} = H_0/100$\,km/sec/Mpc, results in the
{\em improved $Om$ diagnostic} \cite{omh2}
\ber
Omh^2(z_2;z_1) &=& \left\lbrack \frac{h^2(z_2)-h^2(z_1)}{(1+z_2)^3 - (1+z_1)^3}\right\rbrack h^2_{100}\nonumber\\
&=& \frac{h_{100}^2(z_2) - h_{100}^2(z_1)}{(1+z_2)^3 - (1+z_1)^3}\,, 
\eer
where $h_{100}(z) = H(z)/100$km/sec/Mpc.
A significant advantage of $Omh^2$ is that for $\Lambda$CDM:
\beq
Omh^2 = \Omega_{0m} h_{100}^2\,.
\label{eq:concordance}
\eeq
Since observations of the CMB inform us that \cite{planck} $\Omega_{0m} h_{100}^2 = 0.1426 \pm 0.0025$,
 it follows that for $\Lambda$CDM \cite{omh2}
\beq
Omh^2(z_2;z_1) = 0.1426 \pm 0.0025~.
\eeq
Therefore a departure of $Omh^2$ from this value would signal that DE {\em is not}
 $\Lambda$.


A recent reconstruction of the $Om$ diagnostic from DESI-BAO data suggests that cosmic acceleration may be slowing down \cite{calderon-DESI}.
In this context it is interesting to note that several years prior to the release of DESI data, Alexei Starobinsky drew attention to an interesting new model in which DE is metastable and can decay into dark matter (or radiation) via \cite{star-metastable1}
\ber
{\dot \rho}_{_{\rm DE}} &=& - \Gamma \rho_{_{\rm DE}}\nonumber\\
{\dot\rho}_m + 3H\rho_m&=& \Gamma\rho_{_{\rm DE}}\,.
\label{eq:metastable}
\eer

The decay of DE is based on its intrinsic properties and does not depend upon the external environment (expansion rate, etc.), so in this sense (\ref{eq:metastable}) resembles the {\em radioactive decay} of a substance.
Eq. (\ref{eq:metastable}) leads to transient acceleration which may be in better agreement with data than $\Lambda$CDM \cite{calderon-DESI,star-metastable2}; also see \cite{arman-slowing}.


\section{Discussion}
\label{sec:dis}
Alexei Starobinsky was not only an exceptional scientist, but also a very kind and generous human being, a wonderful PhD guide and a team player par excellence. In this latter role he helped mentor several generations of cosmologists including myself,
Lev Kofman,
and numerous others, many of whom have contributed articles to this memorial volume.
During his visits to India, Alexei interacted with many young researchers including Tarun Souradeep, Dipak Munshi, Tarun Saini, Ujjaini Alam, Minu Joy, Arman Shafieloo and Swagat Mishra.  Many of these scientists have gone on to become leaders in the international scientific community and have mentored a whole new generation of young students.  I therefore firmly believe that although Alexei Starobinsky is no longer present amongst us, the legacy of the Zeldovich-Starobinsky school of cosmology will endure for many generations to come.

\section{Acknowledgements}
I thank Swagat Mishra for a careful reading of the manuscript and for his help with the figures.

\newpage

\section{Down memory lane}
\label{sec:picure}

\begin{figure}[H]
\centering
\includegraphics[width=0.8\textwidth]{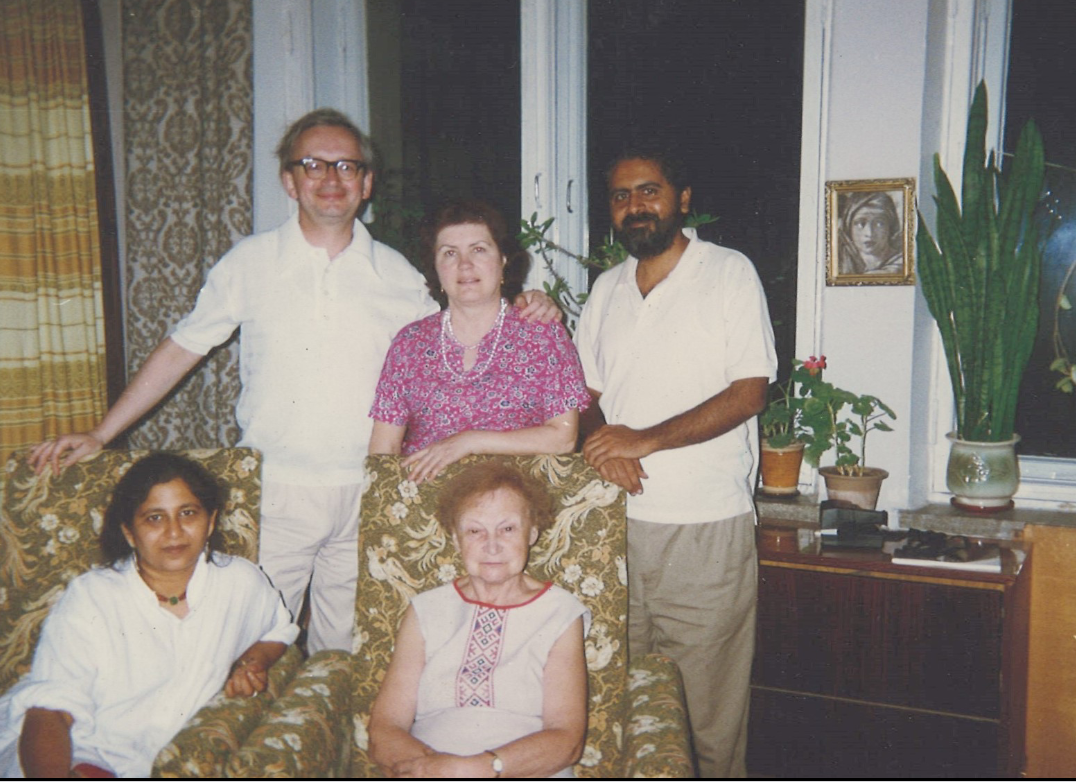}
\caption{ With Alexei at his home in Moscow, clockwise: Alexei, Lyudmila, myself, Alexei's mother and Rohini.
}
\label{fig:Alexei_home}
\end{figure}
\begin{figure}[H]
\centering
\includegraphics[width=0.8\textwidth]{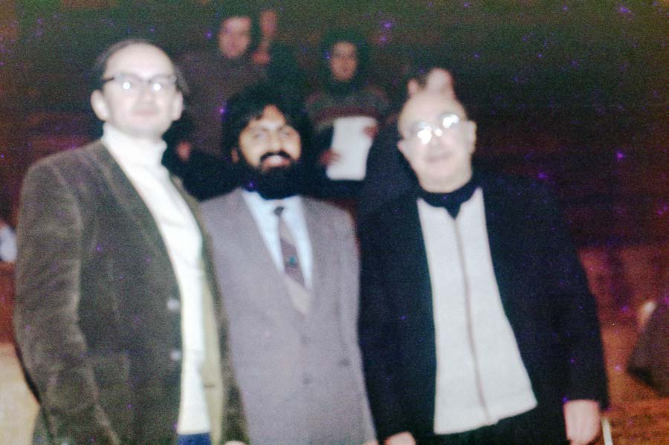}
\caption{With Alexei Starobinsky and Yakov Zeldovich during of my PhD defence, Moscow 1985.
}
\label{fig:my_PhD_defence}
\end{figure}

\begin{figure}[H]
\centering
\includegraphics[width=0.8\textwidth]{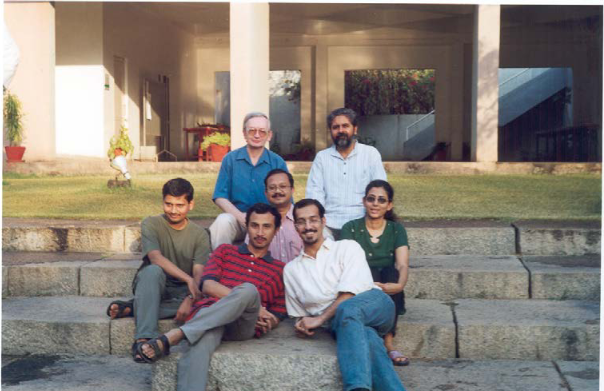}
\caption{Alexei at IUCAA. {\em Top row}: Alexei and myself,
{\em middle row}: Jatush Sheth, Tarun Souradeep and Ujjaini Alam,
{\em bottom row}: Sanjit Mitra and Amir Hajian.
}
\label{fig:Alexei_iucaa}
\end{figure}
\begin{figure}[H]
\centering
\includegraphics[width=0.8\textwidth]{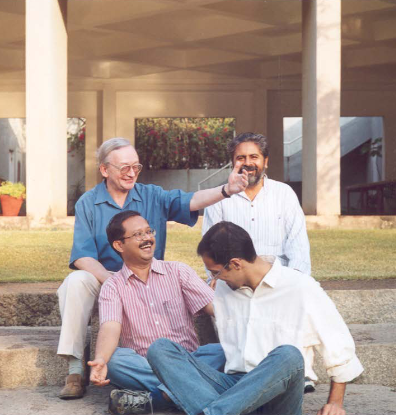}
\caption{Alexei sharing a joke with Tarun Souradeep, Amir Hajian and myself.
}
\label{fig:Alexei_iucaa1}
\end{figure}

\begin{figure}[H]
\centering
\includegraphics[width=0.8\textwidth]{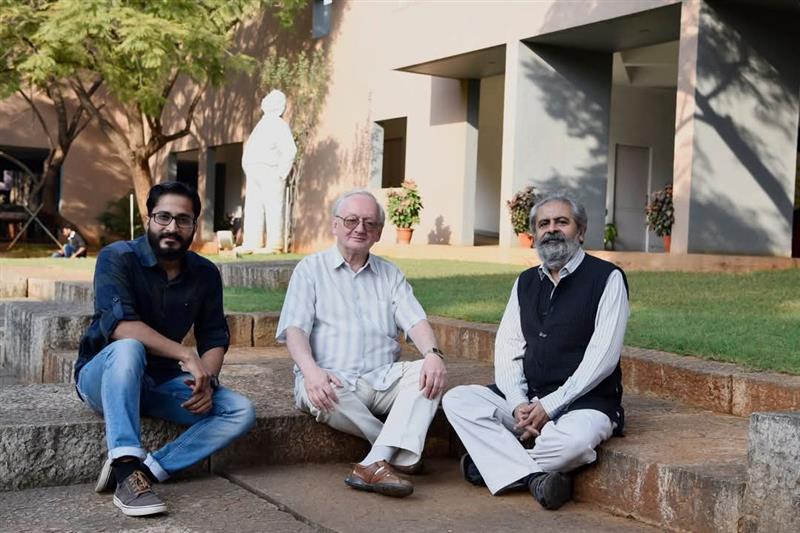}
\caption{Alexei flanked by Swagat Mishra and myself, during Alexei's last visit to IUCAA in 2019.
}
\label{fig:Alexei_iucaa2}
\end{figure}


\begin{thebibliography}{99}

\bibitem{zeldovich}
V. Sahni, {\em ``Ya. B. Zeldovich (1914-1987): Chemist, Nuclear Physicist, Cosmologist''},
IAU Symp. 308, 25-31 (2014) [arXiv:1310.8048].



\bibitem{zeld-star71}
Ya.B. Zeldovich and A.A. Starobinsky,
Sov.Phys.JETP 34, 1159-1166 (1972), Zh.Eksp.Teor.Fiz. 61, 2161-2175 (1971).

\bibitem{lukash-star74}
V.N. Lukash and A.A. Starobinsky,
Sov.Phys.JETP 39742 (1974).

\bibitem{collins-hawking}
C.B. Collins and S.W. Hawking,
Astrophysical Journal, Vol. 180,  317-334 (1973).

\bibitem{star83}
A.A. Starobinsky JETP Lett. 37, 66-69 (1983).

\bibitem{wald83}
R.M. Wald,  Phys.Rev.D 28, 2118-2120 (1983).

\bibitem{moss-sahni}
I.G. Moss and V. Sahni Phys.Lett.B 178, 159-162 (1986).

\bibitem{star73}
A.A. Starobinsky, Sov.Phys. JETP 37, 1 (1973).

\bibitem{hawking74}
S.W. Hawking, Nature, 248, Issue 5443, pp. 30-31 (1974).

\bibitem{hawking_book}
S.W. Hawking, {\em A Brief History of Time}, Bantam Dell Publishing Group, UK (1988).

\bibitem{star79}
 A.A. Starobinsky, JETP Lett.,
30, 682 (1979)

\bibitem{star80}
A. A. Starobinsky, \plb {\bf 91}, 99-102  (1980).

\bibitem{star82}
A.A. Starobinsky,  \plb 117, 175-178 (1982).



\bibitem{star85}
A.A. Starobinsky, Soviet Astronomy Letters, 11, 323 (1985)

\bibitem{star86}
A.A. Starobinsky, Lecture Notes in Physics 246, 107-126 (1986).

\bibitem{star-yokoyama}
A. A. Starobinsky and J. Yokoyama, Phys. Rev. D 50, 6357-6368 (1994)
[arXiv:astro-ph/9407016].

\bibitem{kofman94}
L. A. Kofman, A. D. Linde, and A. A. Starobinsky, Phys. Rev.
Lett. 73, 3195 (1994).

\bibitem{kofman96}
L. A. Kofman, A. D. Linde, and A. A. Starobinsky, Phys. Rev.
Lett. 76, 1011 (1996).

\bibitem{kofman96a}
L.  Kofman,  in
Relativistic  Astrophysics:  A  Conference  in
Honor of Igor Novikov’s 60th Birthday
, Proceedings, Copen-
hagen,  Denmark,  1996,  edited  by  B.  Jones  and  D.  Marcovic
~
Cambridge  University  Press,  Cambridge,  England
[arXiv:astro-ph/9605155].

\bibitem{kofman97}
L. A. Kofman, A. D. Linde, and A. A. Starobinsky, Phys. Rev. D56, 3258 (1997).

\bibitem{yuri95}
Y. Shtanov, J. Traschen, and R. Brandenberger, Phys. Rev. D
51,  5438 (1995);
also see  Y.  Shtanov,  Ukr.  Fiz.  Zh.
38,  1425 (1993)
(in Russian)

\bibitem{gliner}
E.B. Gliner, Sov. Phys. JETP 22, 378 (1966);
E.B. Gliner, Sov. Phys. Dokl. 15, 559 (1970);
E.B. Gliner and I.G. Dymnikova, Sov.
Astron. Lett. 1, 93 (1975);
I.G. Dymnikova, JETP
63, 1111–1115 (1986);
For a nice article on Gliner's important work and difficult life see
D. Yakovlev and A. Kaminker
{\em Nearly Forgotten Cosmological Concept of E. B. Gliner},
arXiv:2301.13150.

\bibitem{gliner1}
A.A. Starobinsky and Ya.B. Zeldovich, Spontaneous creation of the Universe; My Universe, Selected Reviews; Zeldovich, Y.B., Sazhin, M.V.,
Eds.; Harwood Academic Publishers: Chur, Switzerland, 97–134 (1992).

\bibitem{guth81}
A. H. Guth, \prd {\bf 23}, 347 (1981).

\bibitem{linde82}
A. D. Linde, \plb {\bf 108}, 389 (1982).

\bibitem{alstein82}
A. Albrecht and P.J. Steinhardt, \prl {\bf 48}, 1220 (1982).

\bibitem{allen88}
B. Allen, Phys. Rev.
D37, 2078 (1988).

\bibitem{sahni90}
V. Sahni, Phys. Rev.
D42, 453 (1990).

\bibitem{giovani}
 M. Giovannini, Phys. Rev. D58, 083504 (1998); Phys.
Rev. D60, 123511 (1999).

\bibitem{sami2002}
V. Sahni,  M. Sami and T. Souradeep,  Phys. Rev.
D65, 023518 (2002) [arXiv:gr-qc/0105121].



\bibitem{dany_18}
C.~Caprini and D.~G.~Figueroa,
Class. Quant. Grav. 35, no.16, 163001 (2018)
[arXiv:1801.04268].

\bibitem{dany_19}
D.~G.~Figueroa and E.~H.~Tanin,
JCAP \textbf{08} (2019), 011
[arXiv:1905.11960].

\bibitem{swagat}
S.~S.~Mishra, V.~Sahni and A.~A.~Starobinsky,
    JCAP 05 075 (2021)
[arXiv:2101.00271].

\bibitem{birrel-davies}
N.D. Birrel and P.C.W. Davies, {\em Quantum Fields in Curved Space}
(Cambridge Monographs on Mathematical Physics) (1984).



\bibitem{sahni-habib}
V. Sahni and S. Habib,
    Phys.Rev.Lett. 81, 1766-1769 (1998)
[arXiv: hep-ph/9808204].


 
\bibitem{parker68}
L. Parker, Phys. Rev. Lett. 21, 562 (1968)

\bibitem{grish75}
 L.P. Grishchuk, Sov. Phys. JETP
40
, 409 (1975).


\bibitem{linde86}
A.D. Linde, Physics Letters B. 175 (4), 395–400 (1986).

\bibitem{linde08}
A.D. Linde, "Inflationary cosmology." Springer Berlin Heidelberg, 1–54 (2008).

\bibitem{woodard}
R.P. Woodard, {\em Recent developments in stochastic inflation}, arXiv:2501.15843.

\bibitem{vilenkin-ford}
A. Vilenkin and L.H. Ford, Phys. Rev. D 26, 1231 (1982).



\bibitem{whitt84}
B. Whitt, Phys. Lett. B145, 176 (1984).

\bibitem{maeda88}
K. I. Maeda, \prd {\bf 37}, 858 (1988).

\bibitem{gorbunov13}
D. S. Gorbunov and A. G. Panin, \plb {\bf 743}, 79-81 (2015)  [arXiv:1412.3407].

\bibitem{mishra-inflation}
S.~S.~Mishra, V.~Sahni and A.~A.~Toporensky,
    Phys.Rev.D 98, 8, 083538 (2018)
[arXiv:1801.04948].

\bibitem{dowker}
J.S. Dowker and R. Critchley, Phys. Rev. D 13, 3224 (1976).

\bibitem{sahni-kofman}
V. Sahni and L.A. Kofman, Phys. Lett. A 117, 275 (1986).


\bibitem{linde90}
A.D. Linde, {\em Particle Physics and Inflationary Cosmology}, Harwood, Chur, Switzerland (1990).





 




\bibitem{Kallosh-Linde}
R. Kallosh and A.D.  Linde, JCAP07 (2013) 002 [arXiv:1306.5220].












\bibitem{BICEP-Keck}
P. A. R. Ade et al.
Phys. Rev. Lett. 127, 151301 (2021) [arXiv:2110.00483].



\bibitem{BBO}
W.~T.~Ni,
Int. J. Mod. Phys. D 25, no.14, 1630001 (2016)
[arXiv:1610.01148].



\bibitem{DE}
S.~J.~Perlmutter, et al., Nature {\bf 391}, 51 (1998);
\apj {\bf 517}, 565 (1999) [arXiv: astro-ph/9812133].
A.~G.~Riess, et al., Astron. J. {\bf 116}, 1009 (1998)
[arXiv: astro-ph/9805201].


\bibitem{zel68} Ya.B. Zeldovich, Sov. Phys. JETP
Lett. 6, 316 (1967); Sov.Phys. - Uspekhi, 11, 381 (1968).

\bibitem{kofman-star-CMB}
L.A. Kofman and A.A. Starobinsky,
Soviet Astronomy Letters, 11, 271-274 (1985).

\bibitem{loiter}
V. Sahni, H. Feldman and A. Stebbins,
    Astrophys.J. 385, 1-8 (1992).
\bibitem{peebles-ratra}
B. Ratra,  and P.J.E. Peebles, Phys. Rev. D, 37, 3406 (1988).

\bibitem{ostriker-stein}
J.P. Ostriker, and P.J. Steinhardt, Nature 377, 600 (1995).

\bibitem{de_rev}
V. Sahni and A. A. Starobinsky, Int. J. Mod. Phys. D 9, 373 (2000);
P. J E. Peebles and B. Ratra, Rev. Mod. Phys. 75, 559 (2003);
T. Padmanabhan, Phys. Rep. 380, 235 (2003);
V. Sahni, Lect. Notes Phys. 653, 141 (2004);
V. Sahni, {{\tt astro-ph/0502032}};
E. J. Copeland, M. Sami and S. Tsujikawa, Int. J. Mod. Phys. D 15, 1753 (2006);
J. A. Frieman, M. S. Turner and D. Huterer, Ann. Rev. Astron. Astroph. 46, 385 (2008);
S. Tsujikawa, "Dark Matter and Dark Energy: a Challenge for the 21st Century", arXiv: 1004.1493;
S. Nojiri and S. D. Odintsov, Phys.Rept. 505, 59 (2011);
T. Clifton, P. G. Ferreira, A. Padilla and C. Skordis, Phys.Rept.  513, 1 (2012).

\bibitem{saini00}
T.D. Saini, S. Raychaudhury, V. Sahni and A.A. Starobinsky,
    Phys.Rev.Lett. 85, 1162-1165 (2000)
[arXiv:astro-ph/9910231].

 \bibitem{ss06}
V. Sahni and A.A. Starobinsky,   Int.J.Mod.Phys. D15,  2105-2132 (2006)
[arXiv:astro-ph/0610026].

\bibitem{cpl}
M. Chevallier and D. Polarski, IJMP D 10, 213 (2001);
E. V. Linder, \prl 90, 091301 (2003).

\bibitem{coles}
P. Coles and F. Lucchin, 1995, "Cosmology, The origin and evolution of large
scale structure", John Wiley \& sons.

\bibitem{sssa02}
V. {Sahni}, T.D. {Saini}, A.A. {Starobinsky} and U. {Alam},
\jetpl {\bf 77} 201 (2003) [arXiv:astro-ph/0201498].


\bibitem{arman}
A. Shafieloo, U. Alam, V. Sahni and A.A. Starobinsky,
    Mon.Not.Roy.Astron.Soc. 366, 1081-1095 (2006)
[arXiv:astro-ph/0505329].

\bibitem{arman-Gauss}
A. Shafieloo, A.G. Kim and E.V. Linder,
    Phys.Rev.D 85, 123530 (2012)
[arXiv:1204.2272].

\bibitem{alam}
 U. Alam, V. Sahni, T.D. Saini and A.A. Starobinsky,   Mon.Not.Roy.Astron.Soc. 344, 1057 (2003).
[arXiv: astro-ph/0303009].



\bibitem{maryam}
 M. Arabsalmani and V. Sahni, Phys.Rev.D 83, 043501 (2011)
[arXiv:1101.3436].

\bibitem{Om}
V. Sahni, A. Shafieloo and A.A. Starobinsky, Phys.Rev.D 78, 103502 (2008)
[arXiv:0807.3548].


\bibitem{om3}
A. Shafieloo, V. Sahni and A.A. Starobinsky,
    Phys.Rev.D 86, 103527 (2012)
[arXiv:1205.2870].

\bibitem{omh2}
V. Sahni, A. Shafieloo and A.A. Starobinsky,
    Astrophys.J.Lett. 793 (2014) 2, L40 (2014)
[arXiv:1406.2209].

\bibitem{planck}
P. Ade,  {et~al.}, 2013, {\em Planck 2013 results. XVI. Cosmological parameters} A\&A 571, A16 (2014) [arXiv:1303.5076].

\bibitem{calderon-DESI}
R. Calderon, et al. JCAP10(2024)048
[arXiv:2405.04216].

\bibitem{star-metastable1}
A. Shafieloo, et al. Mon.Not.Roy.Astron.Soc. 473, 2, 2760-2770 (2018)
[arXiv:1610.05192].

\bibitem{star-metastable2}
X. Li, et al. Astrophys.J. 887 153 (2019); [arXiv:1904.03790].

\bibitem{arman-slowing}
U. Alam, V. Sahni and A.A. Starobinsky,
JCAP 04 (2003) 002 [arXiv:astro-ph/0302302];
A. Shafieloo, V. Sahni and A.A. Starobinsky,
Phys.Rev.D 80, 101301 (2009)
[arXiv:0903.5141].

\end{thebibliography}
\end{document}